# Subjectifying objectivity: Delineating tastes in theoretical quantum gravity research

Thomas Krendl Gilbert[1]  and Andrew Loveridge[2]

## Abstract

Research in theoretical quantum gravity has continued expansively even as it has become detached from classic arbiters of research such as direct empirical falsification. This makes it an interesting test case for theories of what motivates and mediates contemporary scientific research and of the nature of scientific objectivity. We conducted 50 semi-structured interviews with researchers in the rival camps of string theory and loop quantum gravity, coded a subset for reoccurring themes, and subjected the resulting data to statistical analysis. To delineate the subjective tastes and the related process of collective consensus-making in contemporary quantum gravity research, we mobilize aspects of Daston and Galison's depiction of the scientific self and its relation to epistemic virtues, Bourdieu's field-centered account of social space, and Kantian notions of aesthetics. We make two key contributions. First, our analysis sheds light on the inner workings of the field by connecting its internal epistemic struggles with approaches to understanding scientific fields. Second, our application of theories of social reproduction to the substance of scientific inquiry allows some substantive generalizations of Daston and Galison's framework.

[1]University of California, Berkeley, USA
[2]University of Texas at Austin, USA

**Correspondence to:**
Thomas Krendl Gilbert, UC Berkeley Graduate Division, Berkeley, CA 94720-5900, USA.
Email: tg340@berkeley.edu

## Introduction

The field of theoretical high energy physics (THEP) has become increasingly unmoored from classic arbiters of scientific research. This includes diminishing hope for further empirical hypothesis testing, difficulties in the formalization of quantum field theory (QFT) that stymie attempts grounded in mathematical rigor, and no alternative conceptual criteria that can be relied upon to produce advances grounded in prior knowledge. Nevertheless, research efforts have expanded even as this state of affairs has left the prestige of this historically high-status field increasingly uncertain.

THEP is therefore an interesting testing ground for what motivates and arbitrates contemporary scientific research. Our aim here is twofold. Empirically, we seek to answer the question: 'What, in the absence of direct experimental research, guides developments in contemporary theoretical physics?' Theoretically, we ask: 'What does such an environment's shifting standards for objectivity tell us about the changing makeup of the scientific self?' To tackle both questions, we apply theories of scientific practice established in the rich literature from science and technology studies (STS) (Bourdieu, 2004; Daston and Galison, 2007; Galison, 1987; Harwood, 1987; Pickering, 1984; Schweber, 1988) to a field that has struggled with this very issue internally (Dawid, 2013; Motl, 2004; Polchinski, 2007; Smolin, 2006; Smolin and Green, 2006; Woit, 2007), in the hope of confirming these theories or prompting refinement and generalization.

We restrict our attention to research on quantum gravity (QG), because it is widely regarded as the most foundational research domain in THEP, and the domain furthest removed from direct experimental tests. In short, it can be treated as a microcosm of the situation in THEP described above.

Our theoretical lens has three parts. First, we portray the perspectives of the individual theorists within the QG community, as they comprise the basic building blocks of our explanation of the field. For this, we build on Daston and Galison's (1992, 2007) characterization of the 'scientific self' and its relationship with various epistemic virtues, objectivity foremost among them. Second, we are interested in mapping out the space of tastes associated with the various 'scientific selves' observed in the field. To justify this focus theoretically, we draw from both Bourdieu's general depiction of 'taste' as an indicator of social position and cultural discrimination (Bourdieu, 2010), as well as the Kantian notion of aesthetics as a subjective cognition of unobserved harmonies that underlie natural phenomena (Kant, 2000). Third, we deploy Bourdieu's specific depiction of the scientific field (Bourdieu, 2004) in order to capture and understand the dynamics of individual researchers' interactions with each other, the wider scientific field, and their own consecrated standards of research practice.

These lenses, once combined, permit a psychologically rich and sociologically capacious understanding of what it feels like and means to undertake research as a member of the QG community. We identified a limited number of recurring tastes, which we coded along 'physical', 'epistemic', and 'professional' axes. From the coded interview data, we are able to show that these tastes are distinct, interrelated, and correlated with allegiance to, and status within, specific quantum gravity camps. They appear sufficient to distinguish members of these camps, and help to visualize the camps' social morphology. Furthermore, this framework may help elucidate how the epistemic virtue of 'objectivity' is achieved in quantum gravity research in the absence of robust empirical or mathematical arbiters, and provides a partial account of why one approach, string theory, has been so dominant relative to its competitors in terms of citations, papers, funding, coveted academic positions, and other status metrics. Our aim is not to provide normative ammunition for either side in the 'string wars' (Brumfiel, 2006), but to provide a strictly empirical account.

We argue that these findings confirm and generalize some of Daston and Galison's historical insights, by aligning them with Bourdieu's account of social reproduction. First, our findings suggest that Daston and Galison's core theses are correct: The notion of the 'scientific self' is not singular and immutable, and in particular has an evolving relationship with the epistemic virtue of 'objectivity'. However, we take our results as evidence that the self varies in social space as well as generational time (Mannheim, 1970). Furthermore, this variation includes non-epistemic attributes, and this variation is important for understanding objectivity in contemporary science. In the modern setting of QG research, working scientists fight for and achieve objectivity by meeting intersubjective standards for consensus. Such a consensus is able to be perceived by practitioners as scientifically valid precisely because of the non-epistemic attributes that allow scientists to act as representatives of physical insights and the professional behaviors that affect the style of research and consensus-making. Consequently, the two QG camps described below comprise distinct transindividuations (Stiegler et al., 2012) of mid-twentieth century particle physics, which have developed distinct epistemic standards reflecting different commitments to and realizations of objectivity.

Objectivity requires entering into a relationship with the specific standards of inquiry by which nature can be legitimately investigated, such that one's findings are able to be understood and respected by anyone likewise capable of fully inhabiting those standards. This is why the couplings between physical, epistemic and professional tastes in an individual theorist may be very important: The way theorists are able to force a reconciliation between how nature should work (childhood passion, pop culture books, legendary exploits of famous physicists), how one could investigate it (graduate training, mentorship, internalized lore of the field), and what others in the community would appreciate (paper citations, rising stars, leading institutes). This reflects how objectivity is bounded, as well as the precise meaning of what it means to belong to a specific camp.

## Distinct roads to quantum gravity

THEP is a subfield of theoretical physics descended from work in the foundations of quantum mechanics, particle physics, and the general theory of relativity. In a relativistic quantum theory like the Standard Model of particle physics, energy is inversely proportional to distance scales so that 'high energy' can be thought of equally as 'short distance'. This is roughly equivalent to 'fundamental' in modern technical parlance, so THEP can be thought of as contemporary fundamental physics.

One major aim of THEP is to solve the 'problem of quantum gravity'. The problem concerns building a unified framework for two of the major pillars of 20th century physics: Einstein's general relativity, which describes the gravitational field and its interaction with matter, and quantum mechanics, which describes nature on the smallest scales. Unlike the electromagnetic field and the fields associated with the nuclear forces, gravity has entirely resisted any straightforward attempt at a quantum mechanical description. Because this problem is of such fundamental significance, and because it is so far removed from experiment, it is regarded as having the highest levels of prestige within the community of theoretical physicists. This subdiscipline, QG, is our subject of interest in this article.

Despite its difficulty, multiple major attempts on the problem have been made. Two approaches, string theory (ST) and loop quantum gravity (LQG), have been especially prominent.

At a high level, ST attempts solution by tweaking known physics – by, for example, replacing particles with strings or membranes and by including additional symmetries until gravitation is able to fit into a fully unified quantum framework. It originally grew out of the so-called S-Matrix approaches to particle physics that were popular in the 1960's and strives to be not just a quantum theory of gravity but a 'grand unified theory' of all forces and particles, a 'theory of everything'. It also has important connections to Yang-Mills non-Abelian gauge theory, which describes the strong nuclear force. Because of this, ST attracts attention from researchers for a number of reasons, not just its gravitational applications. ST researchers outnumber LQG by approximately ten to one worldwide.

Loop quantum gravity comprises a somewhat more circumscribed research program, as it seeks specifically to provide a solution to the problem of quantum gravity, and attempts to rely on the same 'canonical quantization' procedure that was used to build quantum electrodynamics from the corresponding classical (Maxwellian) theory. It does not, at least by design, attempt to unify all particles and forces. It does however also have some mathematical connections to non-Abelian gauge theory, and indeed the 'loop' refers to the 'Wilson loops' used to describe quark-quark interactions in the theory of the strong force.

There is a sense in which the two fields are approaching the problem from opposite ends that correspond to the respective physical theories that each camp works to reconcile and merge. In ST, general relativity is treated as an 'effective field theory' which emerges from, but may or may not have much to do with, a more fundamental description of physics. In LQG the principles and structures of general relativity are implemented from the start as fundamental ingredients. In this sense it is conceivable, if unlikely, that these diverging roads to solving the problem of quantum gravity will ultimately converge.

There are other approaches which have attracted less attention but are equally ambitious. These include twistor theory, causal set theory, and other heterodox approaches. These other approaches are not relevant for us here, but we anticipate future research to explore them further based on interview data collection.

Truly novel empirical data has been absent since the early 1970s, producing a major institutional and epistemic crisis for physics research. We cite as an example the controversy surrounding the experimentally unverified concept of supersymmetry (Lykken and Spiropulu, 2014: 34):

> [R]esults from the first run of the LHC [Large Hadron Collider] have ruled out almost all the best-studied versions of supersymmetry. The negative results are beginning to produce if not a full-blown crisis in particle physics, then at least a widespread panic. The LHC will be starting its next run in early 2015 …. If at the end of that run nothing new shows up, fundamental physics will face a crossroads: either abandon the work of a generation for want of evidence that nature plays by our rules, or press on and hope that an even larger collider will someday, somewhere, find evidence that we were right all along.

Given the lack of experimental confirmation, ST has been criticized as a potentially failed theory of nature, pursued without justification (the same logic applies to LQG). Still, the sheer dominance of ST implies that the behavior of the field is not intellectually random, that there are at least some regulatory mechanisms at work in QG and presumably within ST and LQG respectively. Given the lack of traditional arbiters, these presumably involve the subjective judgements of individual scientists and the collective process of consensus-making. In the words of one pragmatic advocate of ST,

> Overwhelmingly the concentration on string theory is a scientific judgement, made by a very diverse group of theorists. Look at any of the several dozen most well-known string theorists: my own scientific experiences and tastes, both inside and outside string theory, are very different from any of theirs, just as they are from each other. I think of myself as a theoretical physicist first, and cross over the boundaries between string theory and several other fields depending on what looks important and interesting, as do many others. (Polchinski, 2007)

A recent work in the field of QG that has prompted considerable debate is the paper: 'Black holes: Complementarity or firewalls?', known colloquially as the 'AMPS' (after the authors' initials) or 'Firewall' paper (Almheiri et al., 2013). By means of a thought experiment, it argues that black hole physics reveals a fundamental contradiction between four deep principles of physics. One of these principles is the 'equivalence principle' associated with general relativity. Two of them, 'unitarity' and 'effective field theory' are associated with modern particle physics and quantum mechanics. The fourth is a result related to Stephen Hawking's black hole entropy formula. One of the authors is from ST, but the paper does not rely on any string-theoretic assumptions, and so is of interest to both camps.

The solution to this apparent paradox remains unresolved, and so it serves as a useful prompt for revealing the aesthetic and ideological allegiances of theorists ostensibly trying to unify all physics to one theory over the other, as at least one principle must be selected, modified, or even discarded to preserve the others. We make use of this prompt extensively in our interview procedure.

One of our goals is to understand how individual scientists conceptualize themselves as researchers in QG: how they make professional and research decisions, how they prioritize conceptual insights from the physical theories they are trying to unify, and how they make sense of the nature of scientific progress in relation to their own work. We will comment on the extent to which this understanding sheds some light on the regulatory mechanisms in QG research.

## Epistemic regimes, tastes, and social space

The history of THEP since the 1960s has been defined by escalating attempts to replace the rapidly drying well of experimental falsification with distinct heuristics of theoretical judgment. Here we review the terms and growth of this crisis in order to justify our social-scientific theoretical lens, drawing heavily from Pickering's (1984) work, which took mid-twentieth century particle physics as a case for articulating the practical dimensions of both theoretical and empirical work in contemporary science. This account exhaustively described the growth of gauge theory as part of the evolving social and institutional context for particle physics and its close relationship with experiment, specifically its necessity for even grasping the physical reality of quarks. In the 'old physics', experiments led the way by systematically investigating observed phenomena, noting anomalies, and providing results for theorists to pore over. But the mathematization of particle phenomenology and theory development, crystallized in gauge theory, was the setting for a new generation of theorists who internalized the judgment of their experiment-friendly forebears while shedding their institutional and epistemic allegiance to direct falsification and empirical exploration, constituting a 'new physics'.

As synchrotrons ballooned in size and funding for fundamental physics research began to decline in the 1970s, the symbiotic relationship between experts in theory and experiment morphed into a dynamic equilibrium. Inquiry profoundly shifted as quarks were transformed from a heuristic for interpreting experimental outputs to an assumed physical reality worthy of independent collider investigations, with both groups competing to set the terms. This practical attitude towards theory development, consensus-based investigation, and close contact with experiment produced a peculiar ambivalence in disposition. Progress was first evaluated by meeting the demands of experiment through articulating phenomenological constraints on models, but also increasingly by postulating new fundamental (and frequently unobserved) particles to account for those phenomenological constraints, and finally by suggesting yet more fundamental physical properties (such as strings) to explain why those postulates were needed.

In this context, hypothetical properties of nature competed for attention with new experimental findings, phenomenological intuition, and the filling-out of the standard model, as revealed through our interviews with pioneers:

> So certainly one motivation of my old work was to try to get something which satisfies some of the constraints that [Chew] was saying would determine completely the nature of the strong interaction. I think that that was too ambitious. However some elements of this bootstrap [program] came out to be true in string theory or in the newer model. I mean in particular the fact that the high energy behavior of the scattering processes are all determined by some retro theory, Regge trajectories, which at the same time support the elementary particles of the same Regge trajectory control at the same time the high energy behavior and also where your particles or excited states are. And then there was this idea of linear trajectories which you know extended out to infinites and infinite number of particles and infinite number of spins and so on. I think these are two general principles probably to fully determine something. You need to put something else. I mean somehow string theory inputs something on top. It satisfies this principle. But maybe QCD also does the same. (quotes from interviews are edited for readability)

The increasing difficulty of conducting experiments was matched by a growing reliance on conceptual innovation and cross-fertilization between theorists and phenomenologists, who could resolve private ambivalences by drawing on each other's public demonstrations of expertise. Theoretical constructs like quark charm, the S-matrix, and Yang-Mills gauge theory encouraged collaboration between theorists, spectroscopists, and experimenters, and were themselves justified and sustained by these emerging collaborations, as Pickering describes:

> By 1976, charm was central to such a range of practice that arguments could always be marshaled in favour of seeing misfits between prediction and data as important results rather than as serious problems … Such discrepant phenomena were simply seen to call for further theoretical and experimental work, which would itself command the attention of a wide audience. (Pickering, 1984: 272)

New phenomena like charm denoted both physical realities and social arrangements, by defining the rules for reconciling experimental settings – whose number was declining year by year – with trusted conceptual heuristics. The result was that the old common-sense physics, in which experimenters initiated inquiry and provided empirical findings that required phenomenological interpretation, gave way to a gauge-theory world-view whose assumed physical interactions became the testing ground for experimenters to isolate and explore.

Once new physics was dominant and the standard model was complete, high-energy theoretical physicists started looking for a unified gauge theory of the weak, electromagnetic, and strong interactions, referred to as a 'grand unified theory' or GUT. String theory, whose first 'revolution' kicked off in 1984, began as an extension and elaboration of this culture. This is the date of Green-Schwarz anomaly cancelation, the moment string theory fully revealed itself as a candidate theory of quantum gravity. The epistemic shift this demanded from young theorists was profound, as it called for abandoning the context of particle experimentation in favor of exploring mathematical structures directly for physical guidance. The resulting psychological pressure is conveyed in one of our interviews:

> I was … still in this mode where we should be talking about things that should be discovered at the next accelerator. Gravity was something you had to, no matter how fascinating it was, you had to let go of it … [Later in the fall of 1984 I] went in to Witten's office and Witten was all excited because he had just gotten the Green-Schwarz anomaly cancellation paper. … I was still in this kind of, you know, phenomenological mindset. So I said to Edward, 'Well, ok the thing that we should do is figure out what the conditions are for having $N = 1$ Supersymmetry in four dimensions because that's what's going to be relevant phenomenologically because otherwise you don't have chiral fermions and so on.' So we came back after lunch and Emil and I come in to Edward's office … And he worked out the Calabi-Yau condition on the blackboard for us, starting with the ten-dimensional Lagrangian, and Emil, we walked out of the office and Emil said 'I don't know whether I can work here', that was just so, you know, because he did, he WALKED up to the blackboard and started writing and BOOM all these results came out and then he knew about the Friedman-Opiskime paper that pointed out that Kahler manifolds are important in the context of sigma models and so on. So he blew both of us away. And I said oh ok, I have to get interested in this subject in a serious way now.

In our interpretation, Pickering's (1984) distinction between the 'old physics' of experimentally-directed research and the 'new physics' of theory-led experimental confirmation maps onto two of the later epistemic regimes outlined by Daston and Galison.

In their work *Objectivity*, Daston and Galison study the history of the scientific self in the modern era. Their thesis, generally stated, is that the history of science involves a number of epochal shifts between distinctive 'epistemic virtues': truth-to-nature, mechanical objectivity, and trained judgment. These shifts, reflected in the coagulation of novel research settings and social spaces, also reflect dispositional adjustments in the ways scientists think about themselves and their work.

Daston and Galison's (2007) typology of epistemic virtues is meant to constitute successive historical crystallizations of the scientific self:

> Alongside the epistemic virtues of truth-to-nature, mechanical objectivity, and trained judgment emerges a portrait gallery of scientific exempla: the sage, whose well-stocked memory synthesizes a lifetime of experience with skeletons or crystals or seashells into the type of that class of objects; the indefatigable worker, whose strong will turns inward on itself to subdue the self into a passively registering machine; the intuitive expert, who depends on unconscious judgment to organize experience into patterns in the very act of perception. These are exemplary personas, not flesh-and-blood people, and the actual biographies of the scientists who aspired to truth-to-nature, mechanical objectivity, and trained judgment diverge significantly from them. What interests us is precisely the normative force of these historically specific personas, and indeed the very distortions required to squeeze biographies into their mold, to transmute quirky individuals into exemplars. (p. 44)

Most relevant to us here is Daston and Galison's account of the shift in scientific self that took place between early and mid-20th century. Specifically, the early theorists were devoted to a paradigm the authors refer to as 'structural objectivity'. This epistemic stance is characterized by reliance on rigorous mathematical/symbolic representations of reality to ensure objectivity, where objectivity is defined to mean 'the suppression of the self'. Later on, modern particle physics caused a shift to 'trained judgment', where practitioners build a kind of intuitive fidelity to correct identification not through objective standards but by immersive training and judicious application of scientific criteria to vast amounts of possible evidence. In the context of THEP, this style was better suited to studying the massive influx of data from particle accelerators, for which a more 'objective' approach would have been impractical, and also encouraged the formation of 'trading zones' for insights between different branches of physics and mathematics, facilitating the growing autonomy of theoretical work from laboratory settings. It is not our claim that this shift was uniform throughout science, or even throughout THEP, or that in Daston and Galison's account one epistemic paradigm entirely replaces another. However, we take Pickering's work as evidence that at least within THEP 'trained judgment' had largely replaced 'structural objectivity' even before contemporary QG research congealed.

In this paper we test Daston and Galison's assertions, assumptions, and concomitant methodologies by deploying Bourdieusian field theory. Bourdieu (2004) notably agrees that objectivity and subjectivity are closely intertwined. But this is because there is an objective, agonistic social space in which participants must act and register their actions as subjectively meaningful, not because the accrual of scientific standards of objectivity demands a corresponding suppression of subjective feeling or awareness.

Bourdieu provides a meso-sociology of the workings of the field that manifests itself in the personality types of the protagonists.[1] While for Bourdieu this ontology is supported by a rich empirical methodology, including everything from ethnography to multiple correspondence analysis, Daston and Galison instead make claims about historical shifts in epistemic regimes, avoiding strong causal claims about the way field forces 'work on' the people within these communities. We suggest that field theory provides a way to synchronically analyze a phenomenon (epistemic taste) that Daston and Galison treat only diachronically. That is, we expand on Daston and Galison by examining how the scientific self varies in 'space' as well as time. Field theory allows us to both refine and modify Objectivity's assumptions about how a classical epistemic regime (mechanical objectivity) turns into distinctively transposed forms of trained judgment. This transposition has multiple dimensions (ontological, epistemic, professional), but they are nevertheless cleanly separable, and this can be shown statistically as well as qualitatively.

In such an environment, the agonistic churn of the field is regulated by the intellectual comportment of the scientists who work within it, the social structure of the field itself, and the nature of the institutions that support research. In this article, we focus on the first of these factors. We characterize the professional, epistemic and physical tastes that define the types of 'scientific selves' found in the subfield of modern theoretical physics associated specifically with the problem of quantum gravity, with an eye towards how these tastes relate to the evolution of the field.

Finally, in order to delineate these tastes' distinctive flavors, we appeal to Kantian aesthetics (Kant, 2000). For Kant, the primary role of taste in aesthetics is to be teleological: One appreciates a painting as beautiful because it expresses something foundational and perhaps sublime about the human condition or about the essence of nature, which cannot be directly perceived but can be hinted at in works of crafted art. Kant paradoxically describes this faculty as subjective a priori: it is both transcendental (available to us via the mind's innate capabilities, regardless of worldly experiences) and peculiar to oneself (making it tied in some way to one's own perception, perspective, and specific orientation as a rational being). This subjective discernment of order depending on an assumed telos beneath natural phenomena is strikingly similar to the different searches for a grand unified theory of nature in ST and LQG, and helps explain the roles of different taste dimensions in shaping the makeup of each community.

The problem of quantum gravity has been around since the earliest decades of the 20th century but has congealed as an autonomous research program only much more recently, in the 1970's through early 1990's, long after the purported shift towards 'trained judgment'. To be clear, the historical shift to 'trained judgment' from 'structural objectivity' was already achieved through the articulation of the standard model. What we are seeking to add to this story is a much more fine-grained understanding of what happens to the subjective scientific self and the role of objectivity after this transition has already reinvented a broader field of inquiry.

## Interview and coding procedure

We conducted approximately 50 interviews with practicing quantum gravity theorists at various stages of the career ladder, from first- or second-year graduate students to emeritus professors, and in several countries – specifically the United States, Canada, United Kingdom, France, and Germany. The average interview length was 90 minutes, but could range up to three hours with major founding figures of a specific camp. After examining all interviews, discussing key themes, and settling on a coding scheme, 32 of these were selected as the basis for statistical analysis. This subset was chosen for their subjects' embodying distinct stages of the career ladder, the coding consistency made possible via close similarity of substantive topics discussed, and the exhaustiveness of the topics themselves (unlike some interviews whose more narrow focus on technical topics served as grounds for omission). These 32 interviews were coded according to criteria arrived at during the course of overall data collection, while many of the other interviews were conducted on more specific topic areas to fill in our understanding of surrounding context, such as additional perspective on key moments in the history of either field (e.g. the discovery of Green-Schwarz anomaly cancelation).

Interview questions were inspired by our knowledge of prominent papers in quantum gravity from both fields, popular science books and publications by prominent members of both camps (e.g. Dawid, 2013; Smolin, 2006), our hypotheses about the social mechanisms for regulating research (described in the next section), and public testimonies of previously interviewed respondents where available (for example, Joseph Polchinski and Don Marolf's contributions to the firewall argument).

Interviews generally covered three main topic areas, each of which took approximately 30 minutes to discuss. First, participants were asked about their personal path to entering the field (respectively either string theory or loop quantum gravity), including motivating factors, family background, choice of undergraduate and graduate institutions, and (if relevant) their dissertation work. These questions were designed to establish a baseline of familiarity and rapport with the interviewer, and also provided some useful qualitative information and valuable (though subjective) historical context for the evolution of the field.

Second, participants were asked to comment on some recent outstanding work in the wider problem space of quantum gravity, in particular the AMPS paper. This question area was intended to flush out participants' intellectual and affective responses to the 'crystallizing problems' of contemporary work in quantum gravity, and as a proxy for what Bourdieu calls position-taking: the jockeying for recognition and accomplishment that defines agonistic social arenas. For example, the interviewer made note of when respondents would appear physically uncomfortable, pained, stand up or sit down, trace out ideas using a chalkboard or whiteboard, or go for long moments (10 or more seconds) without speaking. These moments were interpreted as expressing the 'field tension' a given participant was feeling as they prepared to take a theoretical stance that would be viewed as controversial by members of their own camp or the wider quantum gravity community.[2] These moments of tension were frequently resolved through a gestural performance by which interviewees consciously adopted the mantle of their chosen community and outlined the warrants behind a given style of work. Beyond Bourdieu's descriptions of the scientific habitus, we felt these moments more specifically reflected Daston and Galison's emphasis on the 'minatory force' of norms of scientific knowledge production, and the consequent eruption of subjectivity in a willed performance of contested assumptions.

Third, participants were asked for their views on the relevant competing community (respectively either loop quantum gravity or string theory), including their evaluation of its merits, major insights, limitations, and future prospects, and were also encouraged to compare these factors against that of their own community. Special attention was paid to those few participants who had worked in both camps over the course of their careers or done work that served useful for both; these are described in more qualitative detail in the next section. Beyond corroborating information expressed in the second topic area, these questions were designed to encourage respondents to speak (or try to speak) with more authority about the state of theoretical physics today and the place of quantum gravity within it, including which community was doing more important work. Participants were thus encouraged to 'speak for' objectivity as they understood it operating in their own community.

Finally, all respondents were asked about the distinctive role of thought experiments in their own community and how this related to the wider role they have played in the history of physics. These questions were included to encourage interviewees to provide further rationale for the physical and epistemic tastes they had aired earlier in the interview, to clear up any confusions about how they viewed their own work in relation to earlier, foundational thought experiments (such as Einstein's elevator and train experiments), and to provide an individual, subjective lens on how interviewees viewed the practices of their own field changing over time. While not our focus here, we anticipate a future article delving more deeply into this topic and what thought experiments in quantum gravity tell us about the social organization of contemporary scientific practice.

We coded interviews on the sentence level according to preference for a certain manner of work. To count as an expression of taste, it must have both subjective appeal and conform to one of the categories outlined in the next section. Mere statements of fact, such as 'The black hole information paradox is unsolved' or 'I spoke to a colleague about this' were taken to be insufficient or applicable to the field as a whole. Expressions of taste that did not conform to our categories of interest were not coded.

Of the 32 interviews an initial subset of 12 were coded by both authors, agreement was found at a level above 90%. Subsequent interviews were coded by one author and subjected to review by the other, and the rate of agreement remained well above 90%.

Our subset of 32 interviews were coded to reflect three core dimensions of expressed taste: physical (denoting a preference for quantum field theory or general relativity), epistemic (denoting a preference for 'visionary' or 'pragmatic' work on quantum gravity), and professional (denoting a preference for 'autonomy' or 'interdependence' when working), as well as charisma (earned either through institutional affiliations or personal proximity to key figures). By tracing how our prepared interview topics fleshed out these taste axes, we generated data to reveal the relation between the inheritance of standards of argument and procedure from mid-century particle physics on the one hand, and the subjectification of these standards through the coagulation of 'scientific taste' on the other. While these tastes are thus historically sedimented and our hypotheses would predict their correlation with specific theoretical camps, they are each simply meant to be opposite categories whose social alignment is not predetermined. Furthermore, we were careful to code statements of taste only when expressing preference, rather than matter-of-fact summaries of how work is typically done.

We note that in principle each category (particularly epistemic tastes) is provisional and could have been coded differently. Subsequent research may interrogate each category in more detail, refine them further, or suggest preferred alternatives to them in light of further data collection and interpretation. Still, we do not treat this scheme as an a priori choice, but rather as part of our substantive contribution: We here confirm the existence of several specific recurring tastes in theoretical physics, and also map out where they are found. We also note that, particularly for senior scientists, these tastes carry a certain moral and minatory weight for how research should be done, not just a personal preference. Much of the care and attention we took in constructing and applying these taste categories went into drawing the fine line between the aesthetics of quantum gravity investigation and the ethics of research standards, while recording that the latter were often codified out of the former's sedimentation after decades of practice.

**Table 1.** The typology of physical tastes.

| Particle physics | General relativity |
|---|---|
| Perspective that emphasizes particle physics and quantum field theory, prioritizing its insights, methods, and/or problems of interest, even within the context of quantum gravity<br>"I loved quantum field theory, I thought that's obviously the right way to think about things" | Perspective that emphasizes gravitation and general relativity, prioritizing its insights, methods, and/or problems of interest, even within the context of quantum gravity<br>"I always approached string theory from a gravitational perspective and liked it as a possible quantum theory of gravity as opposed to most people at that point" |

**Table 2.** The typology of epistemic tastes.

| Pragmatic | Visionary |
|---|---|
| Utility-focused, apodictic, quantitative, rapid-discovery mindset, pluralistic, methodological reflexivity<br>"Even if a little fairy told me that [String Theory] was wrong…it has proven itself over and over again as…useful" | Truth-focused, assertoric, qualitative, emphasizes historical awareness, monistic, empirical<br>"But what interested me in physics was not so much uh 'oh look there is a force that pulls and has that acceleration.' But was um, how is it possible that we develop a set of concepts that match with nature." |

Each taste dimension is briefly described below; anonymity was guaranteed for all respondents as part of the research study design. Tables 1, 2, 3 and 4 present each taste and provide (anonymous) exemplars of each from select interviews.

## Physical tastes

Although the goal of both camps is to unify quantum theory with general relativity, an individual theorist may display a preference for the intuitions, insights or principles of one subject over the other. We found it useful to code subjective allegiance to particle physics versus general relativity, although other categories such as quantum information theory or condensed matter physics could have been included in principle as well as secondary tastes.

**Table 3.** The typology of social tastes.

| Interdependent | Autonomy |
|---|---|
| Preference for working closely with others, and for social configurations that encourage or rely on interaction or cooperation "[Some other researchers] don't pay attention to anything that is going on around them. They work behind closed doors and they form their own research. I have no idea how those people function. I need this constant stimulation." | Preference for working alone or separately from others, and for social configurations that allow a lot of autonomy and freedom from interference "[My dream location for a PhD would be] a place where I could uh have the opportunity to do my own work without uh, without pressure from my supervisor. So I would have to develop my personal, my personal point of view on [my subject]." |

**Table 4.** The typology of charisma.

| Institutional | Personal |
|---|---|
| Revolves around institutions and institutional prestige "Well it was a good place. It was prestigious." | Revolves around individual persons "I was very very impressed by a seminar he gave…he stood up and said he had a dream which was to reduce QCD to an exactly solvable model." |

## Epistemic tastes

Beyond substantive representations of nature, two lines of investigation are present in these camps that are derived from epistemic standards of theory development in the twentieth century. We have respectively labeled these 'pragmatic' and 'visionary' in part because they seem to be inspired by specific representations of scientific progress inherited from the cultures of particle physics and Einsteinian cosmic reflection. These are our most capacious categories, reflecting variegated (though coherent) commitments to a certain style or attitude of working whose relationship with other habituated tastes was of central interest to our research design (see our 'Discussion' section for the further distinction of taste versus style in the context of our empirical findings).

## Professional tastes

Physicists expressed preference for a specific manner of working, which we subdivided along two axes. First, theorists tended to work more comfortably either in groups or in isolation. Rather than a mere workplace platitude, we observed that researchers often assigned great intellectual significance to this as a factor in their own productivity and insights, amounting to a substantive preference for either independence or collaboration. We were careful to code such statements as deliberate (i.e. in the form of an explicit preference) so that tastes could be treated as constitutive of a respondent's affinity for a certain mode of inquiry, rather than an arbitrary practice. In addition, quantum gravity physicists exhibited great concern for status in their respective communities (perhaps compensating for a dearth of empirical evidence or falsification of ideas), but had contradictory and mutually exclusive views on how status is acquired and symbolized.

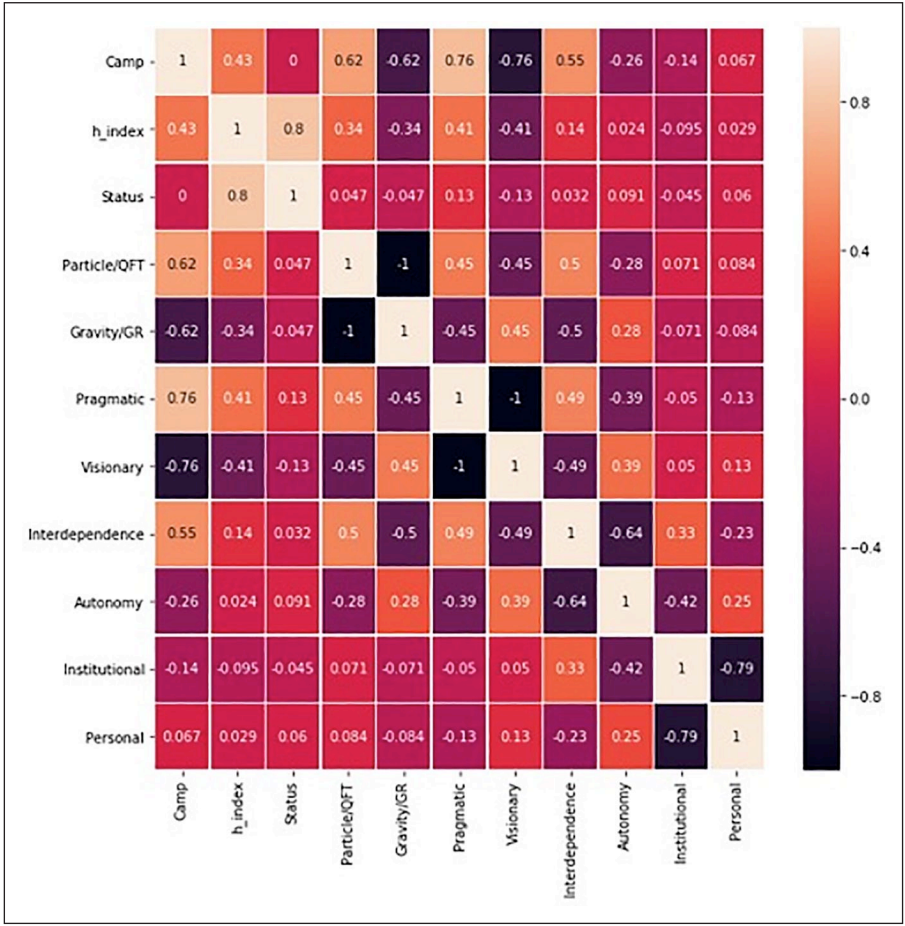

**Figure 1.** Correlations between tastes, status, and camp.

After coding all interview data, we aggregated the counts for all taste dimensions and tabulated these into a scatter plot matrix. Our goal was to uncover correlations between taste dimensions and thereby answer our research questions about the relationship between expressions of theoretical judgment and social position in the field. This is presented in Figure 1.

In order to test our hypotheses about the relationship between 'taste quotient' and status within a specific camp, we bolstered our taste categories with the following information for each theorist: h index (a measure of both the productivity and citation impact of a scientist), string or loop affiliation, professional status (Senior, Mid Level, Junior), physical taste score, epistemic score, professional score.

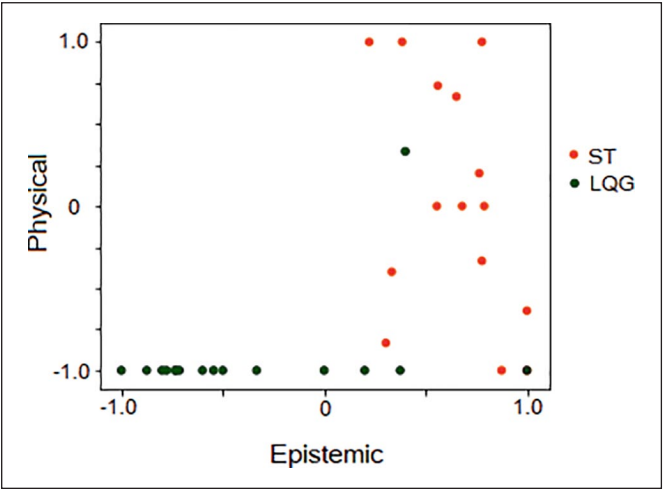

**Figure 2.** Physical and epistemic tastes in ST versus LQG.

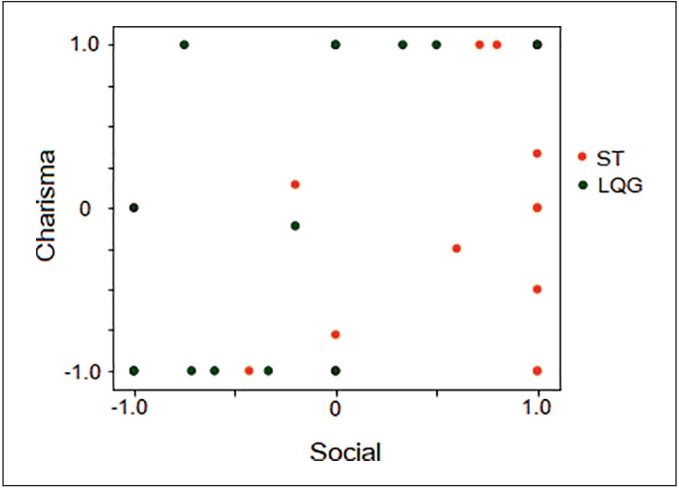

**Figure 3.** Social and charisma tastes in ST versus LQG.

With this information, and armed with our coding scheme, we focus on the following research areas: correlations between the tastes, relationships between the tastes and the camp to which an individual belongs, tastes versus status – for example, how much adherence to the 'correct' tastes is related to status, and how much status is related to having tastes at all – and what qualitative insight these correlations give to certain 'outliers' and singular theorists who have worked in both camps. The resulting correlation graphs are presented in Figures 2, 3, 4, and 5.

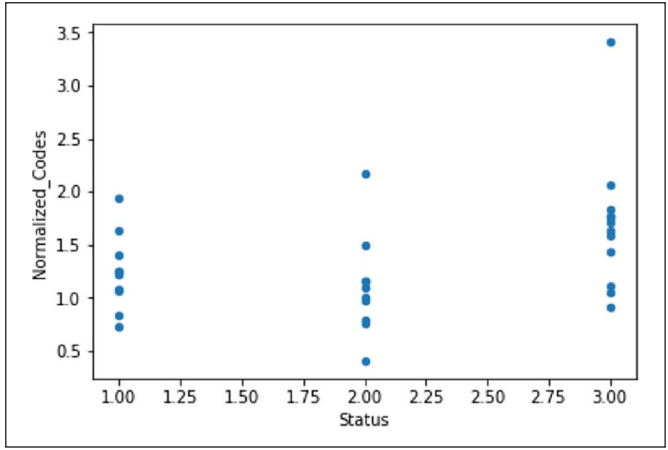

**Figure 4.** Willingness to express taste versus professional status.

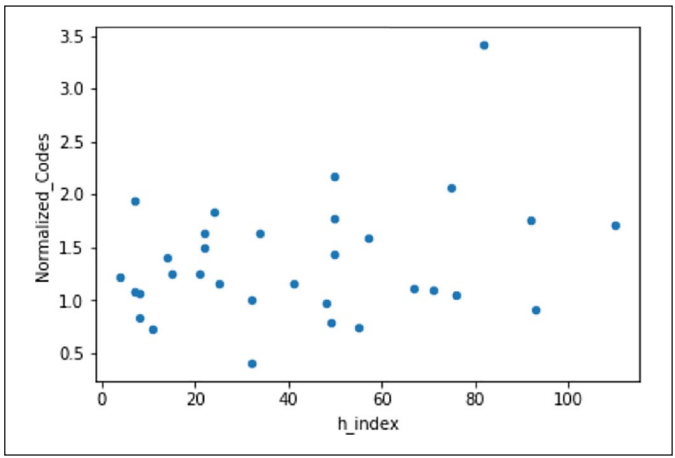

**Figure 5.** Willingness to express taste versus h-index.

## Data analysis

Here we present summary graphs and correlation matrices mapping out the spaces of tastes and their correlators, which we then use to emphasize a few points of interest.

### Correlations between tastes, status, and camp

We make a few key observations from this graph: First, the tastes that we articulated in advance of coding are genuinely opposed to each other, meaning they are opposites to a great extent. Allegiance to one predicts lack of allegiance to the others. This is strikingly so for the epistemic and physical tastes, though less so for the professional tastes. There are also cross-correlations across the axes, for example, GR is correlated with Visionary, etc.

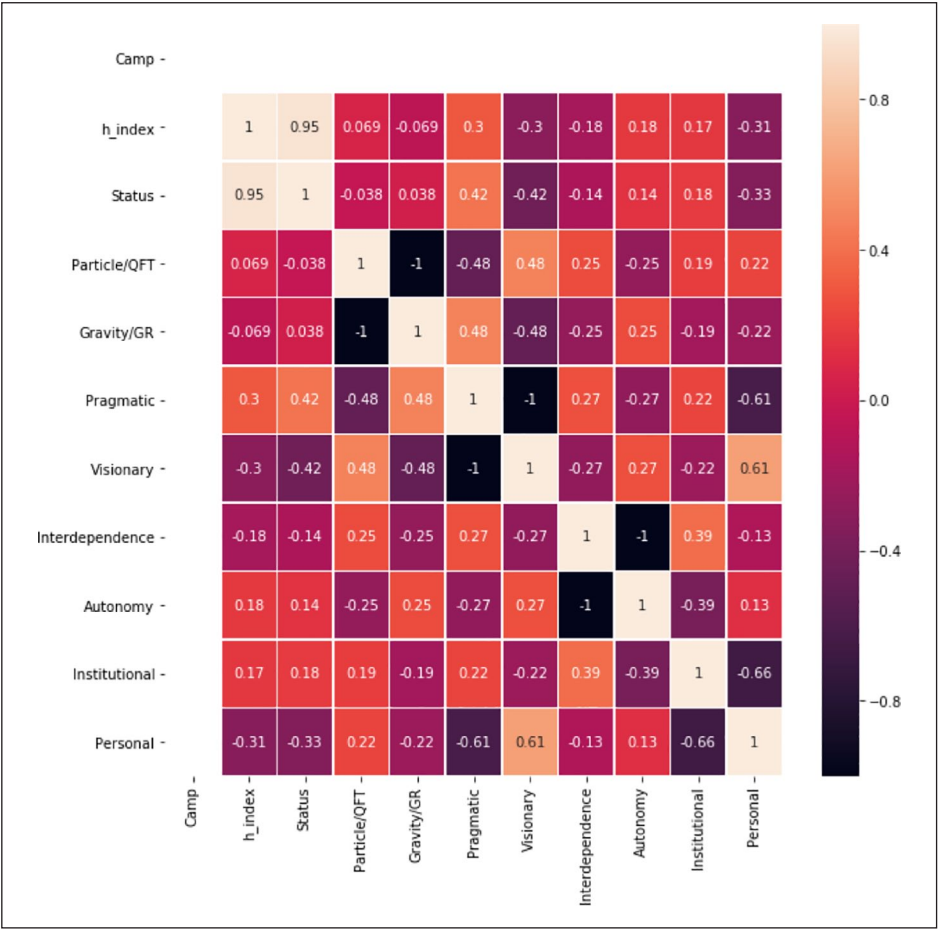

**Figure 6.** String theory's consecration of trained judgment.

Second, by far the strongest correlation with camp is the 'Epistemic' category, followed closely by 'Physical'. Interdependence/Autonomy is also important. The charisma is not very strongly correlated. We note that 'Status' is also correlated with age in our dataset (and in both fields as a whole), so it is possible that older folks just have older 'values' and represent a snapshot of an earlier version of the field. We believe our dataset is large enough that the status correlations nevertheless do reflect the wider social culture of each camp.

## Tastes in ST versus LQG

We here plot the space of tastes for each distinct camp for all individual data points, presented in Figure 6 (ST) and Figure 7 (LQG). The results of the previous graph justify there being only a single axis for each category, which we define as:

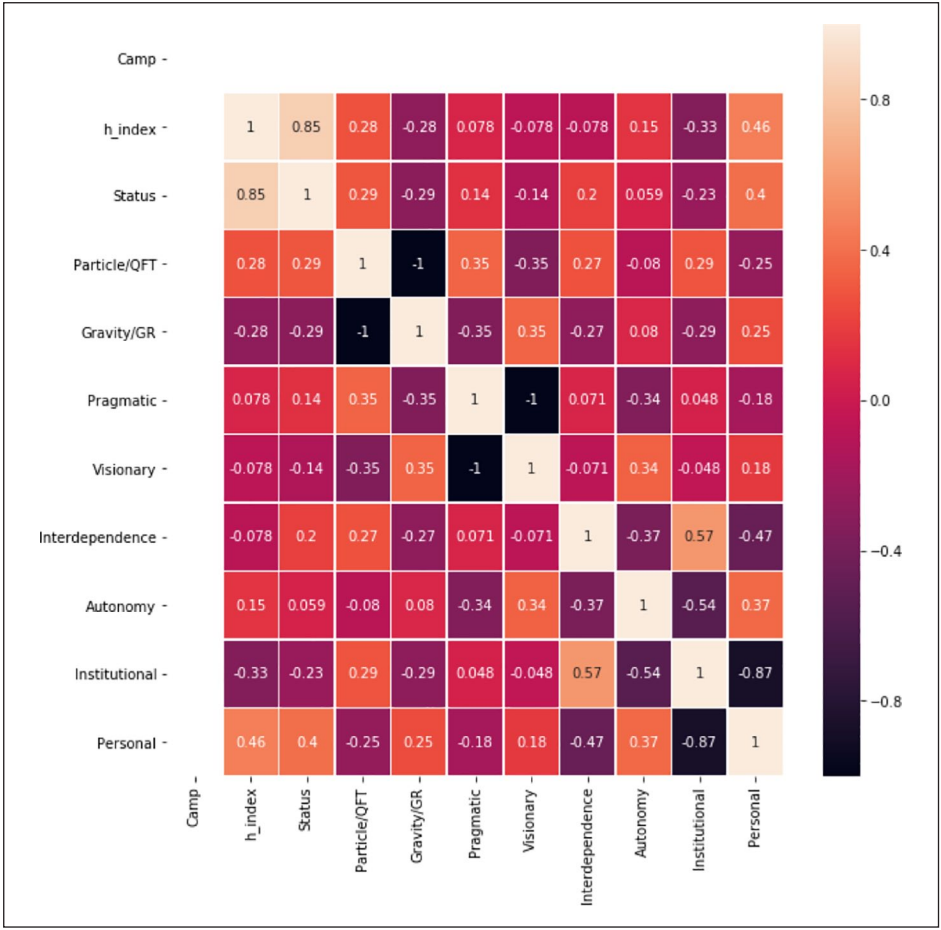

**Figure 7.** Loop quantum gravity's suppression of epistemic virtues.

$$\text{Physical} = (N_{Particle} - N_{GR}) / (N_{particle} + N_{GR})$$
$$\text{Epistemic} = (N_{Prag} - N_{Vis}) / (N_{Prag} + N_{Vis})$$
$$\text{Social} = (N_{Inter} - N_{Auto}) / (N_{Inter} + N_{Auto})$$
$$\text{Charisma} = (N_{Inst} - N_{Pers}) / (N_{Inst} + N_{Pers})$$

Where N is the number of expressions of that taste for that individual.

The ST researchers span the spectrum of physical tastes, and tend to be on the pragmatic side epistemically. However, the LQG researchers almost singularly express the physical taste General Relativity, and cluster on the Visionary end of epistemic tastes but include researchers of all types.

And there are no Visionary Particle Physics researchers in either camp.

The ST researchers tend to cluster in the right half of the second graph (Interdependent spanning the Charisma 4. Spectrum) whereas LQG researchers are of all kinds except the lower right quadrant (Interdependent with Personal Charisma).

We originally suspected that higher status individuals might show more willingness to express taste, but this data suggests that this is either not true or the correlation is very small. It is possible that these tastes are developed extremely early, and determine career trajectory and thus camp selection, rather than the other way around. Thus, people would choose a camp based on their tastes rather than the camp being able to force it on them.

Here, status appears to be related to being pragmatic, has little to do with physical tastes, and is somewhat related to professional tastes. Interestingly, 'Autonomy' and 'Institutional' appear positively connected with status, although they are negatively correlated with each other. This tells us that ST's most eminent figures, despite the field's reputation having grown out of particle theory, are remarkably not defined by their allegiance to a specific physical interpretation of nature.

However, for LQG status does not seem to have much to do with epistemic taste. In fact, status seems to be related to the physical taste 'QFT', which is counterintuitive given that camp's history. Autonomy and personal charisma play a role, which makes sense in this case. It is possible that a physical taste for QFT helps LQG practitioners stand out in their own community and win more citations and recognition from peers.

## Qualitative insights

Members of these communities made statements that suggested that they had normative views about how science should work from a young age, and that this strongly influenced which fields in which they ended up working. It appears that taste categories are highly distinctive, but they seem to deeply depend on each other and grow to support each other over the course of a career. In several interviews, senior representatives gave quotes that combined epistemic, professional and physical categories simultaneously. Several members who had taste combinations that did not reflect the conventional epistemic virtues of their chosen camp suffered career setbacks or major frustrations.

String theory's dominance as a research program seems to stem not just from the material advantages of greater funding or recognition from other theorists, but because of its greater tolerance for disagreement about and pluralism towards the prioritization of variegated physical intuitions. There are prominent qualitative examples of GR-educated theorists who successfully crossed over to ST-style work because their epistemic tastes were highly pragmatic – a disposition that ST actively encourages. These contributors benefited from the pluralism tolerated by the wider ST community, just as their own willingness to commit to a different angle of attack prepared them to contribute to its pluralistic culture, and their unconventional physical tastes could be applied in unique ways to its research program. This is where the insights of field theory, in particular its attention to position-takings, mesh with Daston and Galison's intuition about how epistemic regimes can shift: Being pragmatic makes you objectively more helpful, it makes you subjectively more willing to help, and when combined with unique physical taste produces singular helpfulness in the form of a unique scientific self. As expressed by one interviewee:

> I mean my background was relativity so I always approached string theory from a gravitational perspective and liked it as a possible quantum theory of gravity as opposed to most people at that point who had gotten into it as a particle physicist and were thinking about it more as a sort of unification of all the particles and fields. So I had a slightly different perspective, and so that worked out well I think because I could use some of the tools, and some of the background I had in relativity could then be carried over and get some interesting results in string theory that the other people working in string theory weren't thinking about. So that led to these generalizations of black holes, the black branes that [were] found that later turned out to be important, and other things like that in the '90s.

There is also an important generational dynamic. Younger theorists think and work differently than anyone twenty years their senior, even their own advisors. It might be useful in future work to compare those two against each other to see if this generational shift manifests as specific differences in taste composition, applying the 'problem of generations' (Mannheim, 1970) to a scientific context.

Epistemic priorities are by far and robustly the single most predictive category – predictive in the statistical sense of conditional probability and in the descriptive sense of what it functionally means to belong to a field or how to characterize it, not necessarily in the causal sense of predicting future camp allegience from prior epistemic commitments. This is supported by correlations between taste and camp, from the central role epistemic taste plays in relation to physical and professional ones, and from the way status is defined between camps. Historically, we might describe this in terms of ST inheriting the epistemic stances of particle theory and LQG inheriting those of GR. In fact, this inheritance is more decisive than for those schools' requisite physical tastes, though that is the more common interpretation. Analytically, this seems to corroborate the arguments of Daston and Galison: The evolution of the scientific self is defined first and foremost by epistemic virtues. Furthermore, it casts some doubt on the central role played by symbolic power in Bourdieu's repertoire: Theoretical physicists are most defined not by some specific allegiance to how reality should be represented, but to how such representations must be constructed, maintained, and reasonably altered over time. To be clear, we are not claiming that epistemic tastes are most important for all science, just that this appears to be true for the split between quantum gravity communities.

## Discussion

We organize our takeaway messages in order of our perceptions of their certainty, starting with the conclusions most grounded in our data and ending with more speculative claims.

1. We support and bolster Daston and Galison's pluralistic portrayal of distinctive scientific selves, and how these relate to epistemic priorities. Their insistence on the 'minatory force' of epistemic virtues is profound. They are careful to point out that while their ambition is to uncover the fairly solid regimes through which scientific 'objectivity' has passed, there is not just a wide amount of statistical variance in a given scientist or scientific community's exposure to those regimes,

   but the explanation for these regimes is ultimately rooted in the pressures that face (or do not face) a given scientist in his or her own professional domain. The essence of a scientific theory is inseparable from the wider inertia of scientific practice and, beyond this, novel regimes congeal because what was there before (e.g. the extreme sensitivity to subjective, artistic appreciation of individual specimens or cases that defined truth-to-nature) is exposed to a new kind of 'pressure' or force that has to be rechanneled (e.g. mechanical objectivity is only coherent and even appealing because its rigor and impersonal flavor depends on a willful abnegation of a given scientist's own feelings and subjective impressions while working in a laboratory setting). In the case of our two communities, this pressure took the form of existing standards of argument and disagreement from early- and mid-twentieth century physics that were transposed to a landscape of minimal experimental testing and confirmation.
2. We extend Daston and Galison in two ways:

   (a) We suggest a richer and more contemporaneous depiction of scientific practice defined by a variety of different notions of scientific selves competing for recognition, and which exist at any given moment in time. This takes Daston and Galison's portrayal in a Bourdieusian direction, that is, that the relationship between scientific objectivity and individual subjectivity is deeply tied to the objectivity of social space in which scientists must act and attach meaning to their actions subjectively. Our interview methodology, coding procedure, and statistical analysis seem justified by this perspective. We are making strong causal claims about the way field forces 'work on' and are internalized by the people within these communities, which determines how tastes become correlated with each other and more decisively their specific correlations with status.

   (b) Rival selves include not just epistemic priorities, but professional and ontological tastes as well. We separate the professional, ontological and epistemic dimensions that together constitute a scientist's orientation to the field in question. These other dimensions of taste are not historical or contextual subtleties that should be given less credence, nor are they inherently artificial for bleeding together and working on each other. Rather, we show highly specific cases in which they (a) do affect each other or bleed together, and (b) how these correlations shape or reflect a given scientist's stature in the field, for better or worse. It seems as if reputations may be won and lost based on how well someone can reconcile the 'Neapolitan sundae' of what it means to be in ST or LQG, between these tastes. However, even the recipe for this sundae is comparatively distinct between camps, as in LQG status is won and conferred differently than in ST. And on top of this, LQG is in some ways superficially more similar to 'mechanical objectivity' than trained judgment, which challenges a diachronic depiction of these regimes. This would suggest that the trajectory of epistemic regimes described by Daston and Galison may be arbitrary, that, for example, trained judgment may precede or accumulate differently from the other two depending on its utility for resolving field-specific problems of inquiry.

3. These different types of selves correlate strongly with allegiance to rival paradigms, arguably defining the communities as much as the paradigms themselves. Though we are here more interested in the position-takings than in the positions (i.e. our methodology and analysis leans heavily on exploring the makeup of taste and what makes it tick, rather than the hard contours of the social space in which those tastes congeal), we do provide here a step towards a meso-sociology of the workings of the field as manifested in the personality types of the participants.
4. Our findings offer a clue to how ST may have become so dominant: Its community contains a more diverse mix of scientific tastes, and epistemic attitudes in string theory are more conducive to modern research practices characterized by pluralism.

To expand on this point, it is useful to unpack our interpretation of Bourdieu's understanding of aesthetics. In Bourdieu, 'taste' is partially a function of individual consumption, in the context of individuals' locations and dominant cultures. It has a strong economic flavor, both in that it was ultimately germinated in market settings, and that one must be 'economical' in how one defines one's taste – that is, one can only distribute it in so many different ways. This is what makes taste both objectively real and subjectively meaningful: There are only so many tastes that can define me. Although we have followed the methodological assumptions of field theory in this paper, our interpretation of 'scientific taste' is also close to the teleological spirit of Kantian aesthetics. As we have seen, in both communities one has a 'taste' for a certain manner of investigation, of depictions of the world, and of collaborations with peers because one feels it is more or less likely to result in a final reconciliation with nature. Moreover, one perceives such work as appealing in proportion to the training one has received as a prospective member of a given community. This tension is what animates both communities, and also what helps drive a person originally into one community versus another.

Scientific taste is thus not just a lagging indicator for class or other social inequalities, nor an expression of epistemic fealty, but rather a leading indicator of professional sorting and community belonging. In other words, Kantian aesthetics provides a diachronic treatment of taste unfolding over time that goes beyond and complements Bourdieu's economical-synchronic one, while grounding Daston and Galison's epistemic tastes in the lived professional trajectory of flesh-and-blood scientists. In our case, ST as a field is more inclusive of diverse tastes and is therefore more successful at germinating research in the manner of a 'thought experiment' which relies on aesthetic diversity as a resource, allowing major conceptual innovations like the AMPS paper to take shape and obtain recognition. Consequently, ST has become a highly productive research program, with an enormous market that channels the Bourdieusian economics of taste and a stringent professionalization ladder that shapes its aesthetic purposiveness. Although we are avoiding pseudo-normative evaluations of ST as superior to or more successful than its competitors, we do claim it is 'dominant' epistemically, and not just financially or institutionally, because its aesthetic regime is better aligned with the situational demands and compromises generally faced by late twentieth/early twenty-first century research fields.

5. Objectivity is experienced as a form of apprehensive inquiry into distinct perceptions of the same object, according to some pre-defined domain. As such, we

    zoom out from Merz and Knorr Cetina's (1997) discussion of the determinate path by which a hard problem gets worked on as well as Kennefick's (2000) diagnosis of the theoreticians' regress, favoring a productive tension between stylistic criteria analogous to Kant's distinction between determining judgment and reflective judgment. Mature physical taste is to have a grasp of the poles that mark the boundaries of what it means to work in a given camp (e.g. some triangulation of the principles at stake in the AMPS paper). To have good judgment is to be attuned to this space between physical principles and professional commitments in terms of a spectrum that you can cognitively traverse, and to repeatedly choose specific forms of inquiry that allow you to enter into a closer personal relationship with the object of study which anyone else, likewise attuned, could appreciate. Epistemic taste, that is, how you reliably navigate between the poles of physical intuition and professional responsibility, is the residue of this inquiry.

6. We hypothesize an ontological distinction between 'taste' and 'style' as determinants of scientific membership. In addition to the above discussion points, this is based on an interpretation of our findings in Figure 2 regarding the systematic difference between physical and epistemic measures between communities. In our analysis, taste appears to be predominantly passive and dispositional, functioning as a socialization mechanism into a given community, while style is an active, passionate, and intentional motivating factor for working in and affirming a particular community's approach to inquiry, based on some narrative of how science is supposed to work. This helps contrast our work with recent historiographies of mathematical physics pedagogy (Warwick, 2003) and the adoption of Feynman diagrams (Kaiser, 2009), which according to our definitions hew more closely to 'tastes' rather than 'styles' of scientific work. In the current configuration of quantum gravity physics, the epistemic regimes of *visionary* versus *pragmatic* engagement with theoretical objects appears closer to particular styles of working and thinking given their decisive importance for distinguishing ST and LQG, while the physical and social regimes constitute forms of taste (as they are more widely distributed and commensurate between camps).

## Conclusions

Daston and Galison's account of objectivity can be augmented by accounting for the role of the field in helping to shape, or at least reflect, the dimensions of these priorities in the form of preferences for certain styles of scientific investigation. At least in the case of quantum gravity research, objectivity becomes the responsibility of the wider community rather than any individual; moreover, this responsibility percolates differently across social space. We extend this observation beyond our empirical case, and suggest objectivity in modern science appears to be more a matter of the collective effects of the fields where reputations are established. Many scientists are even aware of their own subjectivity and spin it as a positive through recourse to the need for greater pluralism. This behavior is often a practical necessity.

    Beyond a community's particular thought style (Fleck, 2012), objectivity manifests itself as standards for stylistic consilience. The forms of inquiry (mathematical,

theoretical, phenomenological, experimental) seen as legitimate must be represented in and corroborated through one's investigation. This means that objectivity is a function of the ability to communicate one's investigations to others in a voice that reflects the technique, poise and discursive components necessary for attunement to nature. In speaking to peers, one is able to speak to one's own relationship with the natural world and invite others to share in that precise relationship. As emphasized by Porter (1996), Fuchs (1997), and Ward (1997), the spirit of objectivity is the ability to expose private phenomena to public appraisal and confirmation; but it also materially depends on the social space itself (Bourdieu, 2004). What is remarkable about LQG and ST is that they are distinct modes of such appraisal, expressed in their respective stylistic amalgams of physical, professional and epistemic virtues, and lying in distinct regions of social space. Each community investigates nature according to its own standard; despite cross-community controversies and claims of being not even wrong, neither perceives itself as abnormal in the Kuhnian (Kuhn, 1962) sense. This corroborates Seth's (2007) portrayal of crisis as a discursive construction and commentary on scientific practices that do not conform to one's preferred notion of objectivity.

Thus, corroborating the work of philosophers of science such as Longino (1990, 2002), we present our findings as empirical confirmation of the possibility of objectivity through intersubjectivity, rather than objectivity through the suppression of the self, as an addendum to Daston and Galison's narrative. Interestingly, we found that this mode of investigation existed predominantly, though not exclusively, in ST as compared with LQG. This was perhaps most memorably expressed in an early interview with a researcher who defended the so-called faddish culture of ST, referencing its pragmatic marriage of intersubjective standards of trust with epistemic certainty and an intellectual hierarchy rooted in 'strength of character':

> [I]t's true that there have been periods in the development of this part of theoretical physics … [where] whether or not you're agreeing or not with experiment is uh, there's a meaning to being right or wrong, but it's an internally decided meaning whether something is mathematically consistent whether you've solved a problem or not, because the turnover is actually quicker, in a funny way. It's possible to get little flags, little areas that sorta take off because there are lots of easy calculations to do that no one has done before. And that's another difference between this field and a lot of other fields. Precisely because the number of open questions is small, if a relatively concrete line of exploration is opened up and people focus on it, if you don't want to work on it, you can just go away and come back in a year and know that whatever lesson is there to be learned from it, it'll be picked to the bone, and the good stuff will come out, everything that should be done can be done. You can really trust the confidence and the talent of the people enough such that the only reason a line is not explored is if it's obscure, people haven't thought about it. But if it sorta comes into the limelight for a little while, it's a good thing, because everyone dives in and you sort of know what's easy to do, what'll be cleaned up quickly, and so you just see where the first major roadblocks are along those lines … for better or for worse when you have a field with so many people in it, you know it takes a certain strength of character to attack very hard problems for a very long time and when there's a field filled with interesting problems the sociology of the field more is, that directions open up, people pour in, they completely clean it up, which is really important, which is really great, but then you sort of, it becomes clear if you're not involved in the frenzy, you can just be guaranteed that if you wait six months, it'll sorta clear up what the new vista is.

It is simply not practical for individuals to remain objective in the current era without the help of a field, and resolving whether or not a field remains 'scientific' is not just about mathematical rigor or contact with experiment, but more fundamentally is a matter of the structure of the field as a whole. The scientific self varies in 'space' as well as 'time', and is multifaceted. In addition, certain dimensions of the self (manifesting as taste) may be suppressed or emphasized for the sake of wider community status. In LQG, for example, a personal commitment to autonomy, independent work, and striving for the incorruptible beauty of a geometric cosmos may encourage forms of identity at cross-purposes with the gamesmanship and competitiveness of a modern research environment. ST, by comparison, has a greater tolerance for disagreement and pluralism, but only as long as it remains under the ST umbrella. You can work on ST however you want, but you have to work on ST. This is tolerance in the pragmatic sense we describe in the rest of the paper, not an epistemic virtue in its own right. In other words, ST is highly tolerant of disagreements that are internally productive and aim to build consensus, but not dissent in a stronger, heterodox sense (which by comparison is more a hallmark of LQG practitioners).

> The ST field behaves like an orchestra, where each player contributes uniquely to a coherent composition and must jockey for a limited number of highly respected positions. Its harmony is deeply agonistic. LQG behaves more like an improvised jazz session where everyone is expected to use a smaller range of compatible instruments. From a very young age, theorists learn to develop a taste for either Mozart or Miles Davis, and very rarely both.

At the same time, we observed through our fieldwork and data analysis that quantum gravity research advances largely through collective thought experiments, whose singularity and effervescence reflect the broader field 'coming together' at electric moments that fundamentally realign the research agenda. These thought experiments, which have always been around in some form, are especially interesting because they are conducted and resolved socially. Just as the Large Hadron Collider requires thousands of scientists to complete experiments, ST needs many researchers with different backgrounds to perform and verify its deepest discoveries and convictions. When a social thought experiment is completed successfully, it appears to react back on the structure of the field, reorganizing it around new concepts, status hierarchies, and even the internal personality structures of key scientists (i.e. their tastes etc.). We do not claim to completely understand the dynamics of these thought experiments, but intend to explore those dynamics systematically in a future paper by examining the relation between self and field in the context of specific historical 'thought experiments' conducted in each field. It may be that the process of objectivity through subjectivity is contextualized as a snapshot of an ongoing dynamical process of transindividuation[3] which simultaneously produces the conditions for, and is shaped by, the process of collective consensus-making.

## Acknowledgements

We would like to thank the members of both communities for their willingness to participate in this study. In particular, we wish to thank Nima Arkani-Hamed, Joseph Polchinski, and Carlo Rovelli for their support and insights.

## ORCID iD

Thomas Krendl Gilbert  https://orcid.org/0000-0003-1029-4535

## Notes

1. In this paper, we are more interested in what Bourdieu calls position-takings than in the positions (i.e. our methodology and analysis lean heavily on exploring the makeup of taste and what makes it tick, rather than the hard contours of the social space in which those tastes congeal). In a future paper we intend to explore the latter in more detail by examining the effervescent thought experiments that reset the quantum gravity research agenda over time.
2. We were careful to note and code these moments only for topics on which the interviewee was a recognized expert, so as to exclude mere difficulty in understanding the form of a question or interpreting its content
3. We follow Stiegler et al. (2012) and interpret 'transindividuation' as a description of how the research style(s) of one generation can become crystallized and determine the next generation's grasp of objectivity. This is reflected distinctively in both epistemic standards and the social arenas in which those standards are operative.

## Author biographies

Thomas Krendl Gilbert is an interdisciplinary PhD candidate in Machine Ethics and Epistemology at UC Berkeley. His research interests lie in the philosophy of artificial intelligence, political theory, and the sociology of scientific knowledge.

Andrew Loveridge is a theoretical physicist and education reformer. He's currently an Assistant Professor of Instruction in the Department of Physics at the University of Texas at Austin.